\documentclass[twocolumn,english,aps,superscriptaddress,pre,longbibliography]{revtex4-1}
\usepackage[T1]{fontenc}
\usepackage[latin9]{inputenc}
\usepackage{color}
\usepackage{babel}
\usepackage{amsmath}
\usepackage{amssymb}
\usepackage{graphicx}
\usepackage{esint}
\usepackage[unicode=true,pdfusetitle,
 bookmarks=true,bookmarksnumbered=false,bookmarksopen=false,
 breaklinks=true,pdfborder={0 0 0},pdfborderstyle={},backref=false,colorlinks=true]
 {hyperref}
\hypersetup{
 linkcolor=blue,citecolor=blue,urlcolor=blue}

\makeatletter
\usepackage{pslatex}

\usepackage{color}
\usepackage{babel}
\usepackage[caption=false]{subfig}

\@ifundefined{showcaptionsetup}{}{%
 \PassOptionsToPackage{caption=false}{subfig}}
\usepackage{subfig}
\makeatother

\begin{document}
\title{Contact process with aperiodic temporal disorder}
\author{Ariel Y. O. Fernandes}
\affiliation{Instituto de F\'{i}sica, Universidade de S\~ao Paulo, Rua do Mat\~ao,
1371, 05508-090, S\~ao Paulo, SP, Brazil}
\author{José A. Hoyos}
\affiliation{Instituto de F\'{i}sica de S\~ao Carlos, Universidade de S\~ao
Paulo, C.P. 369, S\~ao Carlos, SP, 13560-970, Brazil}
\affiliation{Max Planck Institute for the Physics of Complex Systems, N\"othnitzer
Str. 38, 01187 Dresden, Germany}
\author{Andr\'e P. Vieira}
\affiliation{Instituto de F\'{i}sica, Universidade de S\~ao Paulo, Rua do Mat\~ao,
1371, 05508-090, S\~ao Paulo, SP, Brazil}
\date{\today}
\begin{abstract}
We investigate the nonequilibrium critical behavior of the contact
process with deterministic aperiodic temporal disorder implemented
by choosing healing or infection rates according to a family of aperiodic
sequences based on the quasiperiodic Fibonacci sequence. This family
allows us to gauge the temporal fluctuations via a wandering exponent
$\omega$ and put our work in the context of the Kinzel--Vojta--Dickman
criterion for the relevance of temporal disorder to the critical behavior
of nonequilibrium models. By means of analytic and numerical calculations,
the generalized criterion is tested in the mean-field limit.
\end{abstract}
\maketitle

\section{Introduction}

Nonequilibrium phase transitions~\citep{Marro2005} offer an interesting
extension of ideas developed in the context of equilibrium critical
phenomena to problems in which time plays a central role. This is
the case of problems featuring absorbing states~\citep{Hinrichsen2000},
such as turbulence in liquid crystals~\citep{Takeuchi2007}, reaction-diffusion
processes~\citep{Alcaraz1994}, and extinction phenomena in biology~\citep{Murray2007}.

A paradigmatic model for these problems is the contact process~\citep{teharris1974},
which can be formulated as describing the dynamics of an epidemics.
The model assumes that individuals fixed at the vertices of a fully
occupied lattice can be either infected or healed. Infected individuals
transmit the infection to its nearest neighbors at a rate $\lambda$,
and become healed at a rate $\mu$. It is now well established that,
for a fixed healing rate $\mu$, there is a critical value $\lambda_{c}$
of the infection rate $\lambda$ separating and active phase ($\lambda>\lambda_{c}$),
in which the epidemics persists indefinitely, from an inactive (absorbing)
phase ($\lambda<\lambda_{c}$), in which the epidemics stops after
a finite time (see, e.g., Refs.~\citealp{Marro2005,Tome2015} and
references therein). 

Remarkably, concepts of equilibrium phase transitions, such as scaling
invariance and universality class, are quite useful to describe the
non-analyticity of this nonequilibrium phase transition. This is because
fluctuations of the order parameter field (the density of infected
individuals $\rho$) are self similar at the transition. This means
that the length $\xi$ and time $\xi_{t}$ scales of this fluctuations
diverge when approaching the critical point. More precisely,
\begin{equation}
\xi_{t}\left(\epsilon\right)\sim\left|\epsilon\right|^{-\nu_{\parallel}},\label{eq:def-nu}
\end{equation}
 where $\epsilon=\lambda_{c}-\lambda$ is the distance from criticality,
and $\nu_{\parallel}$ is the correlation-time critical exponent.
Likewise, $\xi\sim\left|\epsilon\right|^{-\nu_{\perp}}$. The other
critical exponents of interest to our work are the order-parameter
exponent $\beta$, which is defined from 
\begin{equation}
\rho\left(\epsilon\right)\sim\left(-\epsilon\right)^{\beta},\label{eq:def-beta}
\end{equation}
and $\delta$~\citep{faria2008,barghathi2014}, the critical exponent
defining the power-law relaxation of the density at the critical point,
\begin{equation}
\rho_{c}\left(t\right)\sim t^{-\delta}.\label{eq:def-delta}
\end{equation}
In general, we expect $\xi_{t}$ to be related to the time needed
for the asymptotic behavior to set in, so that $\rho\left(\epsilon\right)\sim\xi_{t}^{-\beta/\nu_{\parallel}}\sim\rho_{c}\left(\xi_{t}\right),$
and we conclude that
\begin{equation}
\delta=\frac{\beta}{\nu_{\parallel}}.\label{eq:screl}
\end{equation}
 In the mean-field limit, we have $\nu_{\parallel}=\beta=\delta=1$.
In 1D, these exponents are $\nu_{\parallel}\approx1.73$, $\beta\approx0.28$,
and $\delta\approx0.16$~\citep{Odor2004}.

As in equilibrium, disorder may also have profound effects in nonequilibrium
phase transitions, with the additional aspect that disorder ingredients
may be present over space as well as over time. In ecological models,
for instance, spatial disorder represents the variation of environmental
conditions across the terrain, whereas temporal disorder represents
fluctuations in environmental conditions over time. 

In the contact process, disorder can be implemented by allowing the
rates $\lambda$ and $\mu$ to vary over the sites of the lattice
or over time. The contact process with spatial disorder has been extensively
studied (see, e.g., Refs.~\citealp{Noest1986,Hooyberghs2003,Vojta2005,Vojta2006,Hoyos2008,faria2008,barghathi2014}),
and only recently the effects of temporal disorder has attracted attention~\citep{vazquez-etal-prl11,Vojta2015,barghathi2016,Oliveira2016,Solano2016,Wada2018,Fiore2018,Encinas2021,Wada2021}. 

Prominent among the latter investigations are Refs.~\citealp{Vojta2015,barghathi2016,Wada2018},
which explore how the critical behavior of the contact process is
affected by random temporal disorder, both uncorrelated~\citep{Vojta2015,barghathi2016}
and correlated~\citep{Wada2018} disorder. These works show that
the introduction of temporal disorder induces an infinite-noise critical
point, in which density fluctuations increase without limit with time.
As a result, the ensemble typical and arithmetic averages of the population
density behave quite differently. As time increases, the former becomes
much less than the latter (which is dominated by rare events). 

Disorder is introduced in the model in the following way: we consider
consecutive time intervals of same duration $\Delta t_{n}$. To the
$n$th time interval, we assign an infection $\lambda_{n}$ and healing
$\mu_{n}$ rate which are uniform throughout the lattice. In the random
case, the parameters $\left(\lambda_{n},\mu_{n}\right)$ are chosen
from a probability distribution. Here, in the aperiodic case, $\left(\lambda_{n},\mu_{n}\right)=\left(\lambda_{A},\mu_{A}\right)$
or $\left(\lambda_{B},\mu_{B}\right)$ depending on whether the $n$th
letter of a word $A$ or $B$. This word is obtained employing the
generalized Fibonacci sequence. Starting with the letter $A$, we
apply the inflation rules $A\rightarrow AB^{k}$ and $B\rightarrow A$,
where $B^{k}$ denotes $k$ consecutive letters $B$. For $k=1$,
we recover the original Fibonacci sequence.

Although deterministic, the iterated sequence has no period and is
characterized by intrinsic temporal fluctuations growing as $\sim t^{\omega}$,
with the so-called wandering exponent $\omega$ dependent on the value
of $k$. 

In close analogy with Luck's~\citep{Luck1993b} generalization of
the Harris criterion~\citep{ABHarris1974} for the relevance of spatial
disorder on phase transitions in thermodynamic equilibrium, it is
possible to derive a perturbative criterion for the relevance of aperiodic
temporal disorder on the nonequilibrium case. Near criticality ($\left|\epsilon\right|\ll1$)
and along the characteristic time scale $\xi_{t}$, fluctuations of
$\epsilon$ are of order $\xi_{t}^{\omega}$ {[}see Eq.~(\ref{eq:Gj})
in Appendix~\ref{app:fractions}{]}, so that the corresponding average
fluctuations are 
\begin{equation}
\delta\epsilon\sim\frac{\xi_{t}^{\omega}}{\xi_{t}}\sim\left|\epsilon\right|^{\left(1-\omega\right)\nu_{\parallel}}.
\end{equation}
 Aperiodic temporal disorder is a relevant perturbation to the clean
critical theory if $\delta\epsilon\gg\left|\epsilon\right|$, which,
thus, leads to the criterion
\begin{equation}
\left(1-\omega\right)\nu_{\parallel}<1.\label{eq:relev}
\end{equation}

Notice that for uncorrelated random temporal disorder we have $\omega=\frac{1}{2}$
and the inequality (\ref{eq:relev}) reduces to the temporal version
of the Harris--Luck criterion, $\nu_{\parallel}<2$, formulated by
Kinzel~\citep{Kinzel1985} and by Vojta and Dickman~\citep{Vojta2016}.
In fact, these last authors also investigated the case of correlated
random temporal disorder characterized by a power-law correlations
with an exponent $\gamma$, finding out that in this case the criterion
for instability of the critical behavior in the presence of disorder
becomes $\gamma\nu_{\parallel}<2$. As $\gamma$ is related to the
Hurst exponent by $\gamma=2-2H$~\citep{Wada2018}, the criterion
can also be written as $\left(1-H\right)\nu_{\parallel}<1$. Comparing
with Eq.~(\ref{eq:relev}), we see that, for deterministic aperiodic
temporal disorder, the wandering exponent $\omega$ plays the role
that the Hurst exponent plays for random correlated temporal disorder. 

In this paper, our aim is to test the stability criterion (\ref{eq:relev})
in the mean-field limit of the contact process, which allows for extensive
analytical work to be performed, enabling us to obtain results for
the long-time behavior and the critical exponents of the model. In
Section \ref{sec:Mean-Field-limit}, we sketch the mean-field treatment,
writing a recurrence equation for the density of infected agents at
the beginning of each time interval and analytically determining the
criticality condition. In Section \ref{sec:RG}, we describe a renormalization-group
(RG) treatment that allows us to present analytical results for some
critical exponents, which turn out to depend on $k$. Numerical calculations
needed to extract further information are described in Section \ref{sec:Numerical-results}.
There are also two appendices, describing some technical details.

\section{Mean-Field limit\label{sec:Mean-Field-limit}}

\begin{figure}
\begin{centering}
\subfloat[\label{fig:ilustrarenorma}]{\begin{centering}
\includegraphics[clip,width=0.8\columnwidth]{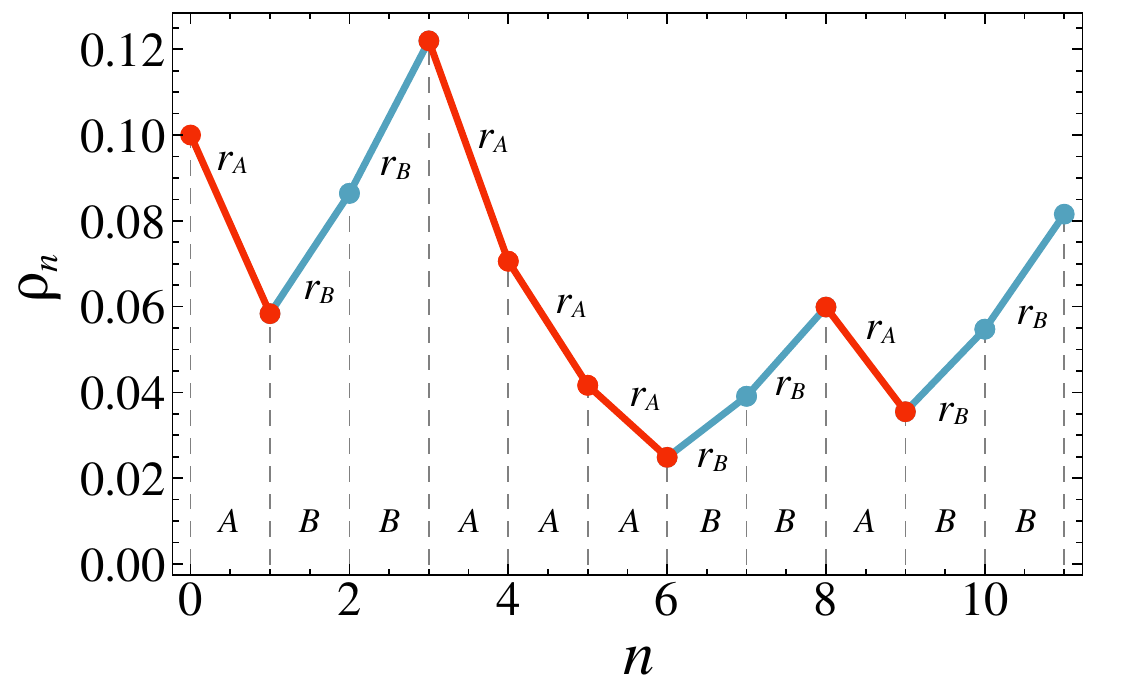}
\par\end{centering}
}
\par\end{centering}
\begin{centering}
\subfloat[]{\begin{centering}
\includegraphics[clip,width=0.8\columnwidth]{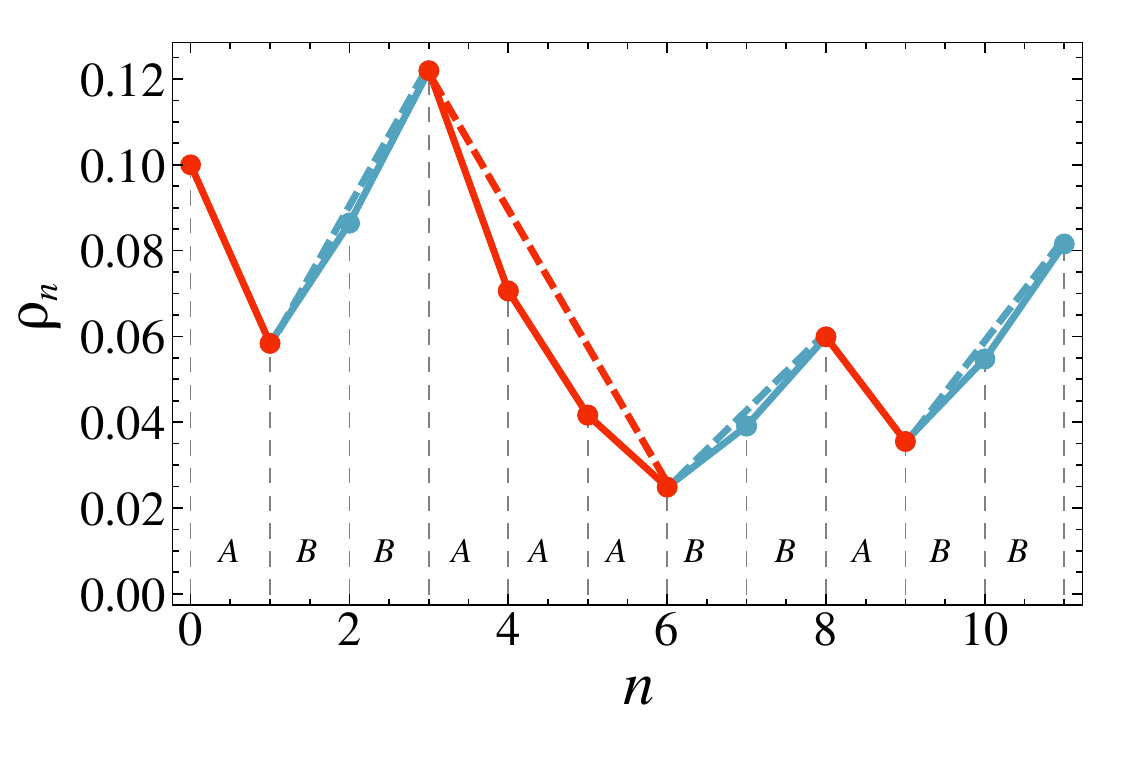}
\par\end{centering}
}
\par\end{centering}
\begin{centering}
\subfloat[\label{fig:ilustrac}]{\begin{centering}
\includegraphics[clip,width=0.8\columnwidth]{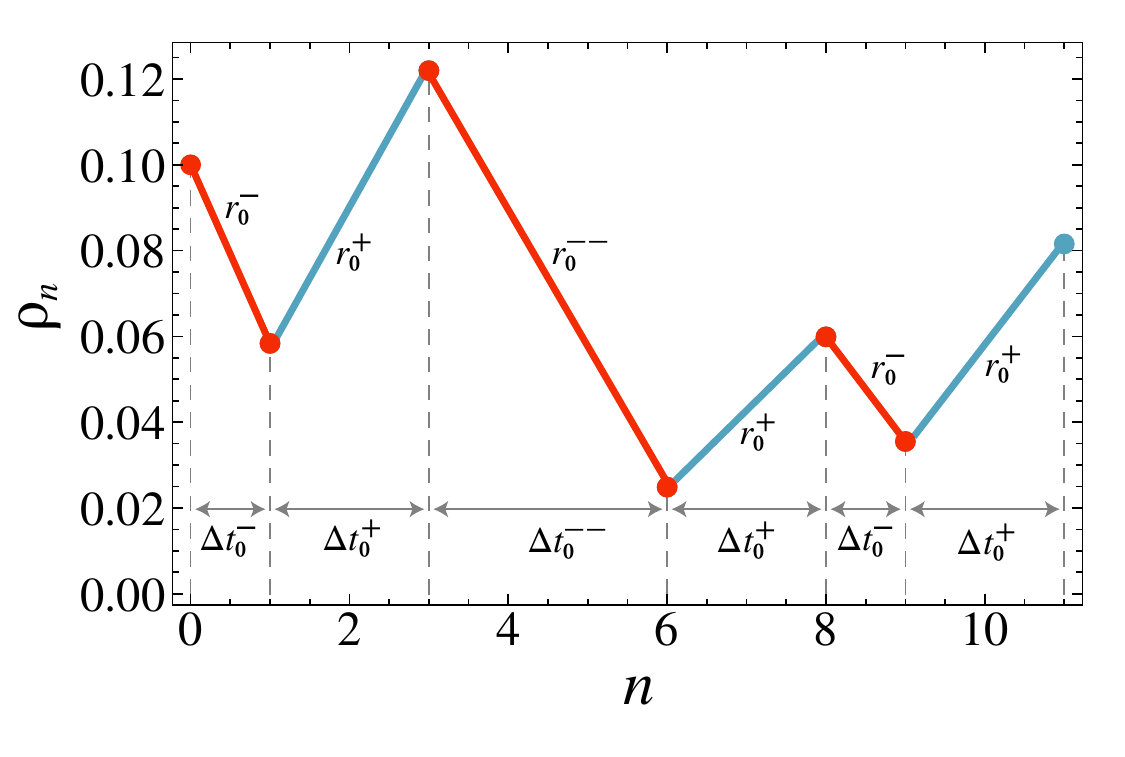}
\par\end{centering}
}
\par\end{centering}
\caption{\label{fig:ilustrarenorm}(a) Plot of $\rho_{n}=\rho\left(t_{n}\right)$
vs. $n$ (with $t_{n}=n$$\Delta t$) for $k=2$ and $\lambda_{A}<\mu<\lambda_{B}$.
(b) Implementing the RG treatment of the dynamics, which consists
in grouping all consecutive time intervals with the same $r_{n}$.
(c) As a result of the grouping, the effective system has parameters
$\tilde{r}$ and $\tilde{\Delta t}$.}
\end{figure}
In this Section we consider the mean-field limit of the contact process
with aperiodic temporal disorder. As $\nu_{\parallel}=1$ in mean-field,
the generalized Harris criterion (\ref{eq:relev}) says that aperiodic
temporal disorder is a relevant perturbation when $\omega>\omega^{*}=0$.
As shown in Appendix~\ref{app:fractions}, $\omega_{k=1}<\omega_{c}$,
$\omega_{k=2}=\omega_{c}$, and $\omega_{k>2}>\omega_{c}$. Thus,
changing $k$ from $1$ to $3$ gives us the rare opportunity to test
the criterion (\ref{eq:relev}) in all possible situations (irrelevant,
marginal, and relevant) by analytical means.

During the time interval between $t_{n-1}$ and $t_{n}=t_{n-1}+\Delta t_{n}$,
the density of active sites can be described by the logistic equation~\citep{Marro2005,Tome2015,Vojta2015,barghathi2016}
\begin{equation}
\frac{d\rho}{dt}=\left(\lambda_{n}-\mu_{n}\right)\rho-\lambda_{n}\rho^{2},\label{eq:logistic}
\end{equation}
in which $\lambda_{n}$ and $\mu_{n}$ are respectively the infection
and healing rates during that time interval, which lasts a time $\Delta t_{n}$.
It is immediate to integrate Eq.~(\ref{eq:logistic}) to obtain,
for $t_{n-1}\leq t\leq t_{n}$,
\begin{equation}
\frac{1}{\rho\left(t\right)}=\frac{e^{\left(\mu_{n}-\lambda_{n}\right)\left(t-t_{n-1}\right)}}{\rho_{n-1}}+\frac{\lambda_{n}\left[e^{\left(\mu_{n}-\lambda_{n}\right)\left(t-t_{n-1}\right)}-1\right]}{\mu_{n}-\lambda_{n}},
\end{equation}
with the notation $\rho_{n}\equiv\rho\left(t_{n}\right)$. Imposing
continuity of $\rho\left(t\right)$ at $t=t_{n}$ leads to the recursion
relation
\begin{equation}
\rho_{n}^{-1}=r_{n}\rho_{n-1}^{-1}+s_{n},\label{eq:logrr}
\end{equation}
with 
\begin{equation}
r_{n}=e^{\left(\mu_{n}-\lambda_{n}\right)\Delta t_{n}}\quad\text{and}\quad s_{n}=\frac{r_{n}-1}{\mu_{n}-\lambda_{n}}\lambda_{n}.\label{eq:ancn}
\end{equation}
Notice that for $\mu_{n}>\lambda_{n}$ we have $r_{n}>1$, while for
$\mu_{n}<\lambda_{n}$ we have $0<r_{n}<1$; as for $s_{n}$, it is
always non-negative. Thus, as expected, $\rho_{n}$ decreases when
$\mu_{n}>\lambda_{n}$ and increase only when $\mu_{n}<\lambda_{n}$,
as illustrated in Fig.~\ref{fig:ilustrarenorma}.

Iterating the recursion relation in Eq.~(\ref{eq:logrr}) yields
$\rho_{n}^{-1}=R_{n}\rho_{0}^{-1}+S_{n},$in which
\begin{equation}
R_{n}=\prod_{i=1}^{n}r_{i}\quad\text{and}\quad S_{n}=s_{n}+\sum_{i=1}^{n-1}s_{i}\prod_{j=i+1}^{n}r_{j}.
\end{equation}
The term $S_{n}$ is responsible for preventing $\rho_{n}$ from becoming
greater than $1$. On the other hand, the fate of the infection when
$\rho_{n}\ll1$ lies essentially on the term $R_{n}$, which, defining
\begin{equation}
\left\langle \lambda\right\rangle _{n}=\frac{1}{t_{n}}\sum_{i=1}^{n}\lambda_{i}\Delta t_{i}\quad\text{and}\quad\left\langle \mu\right\rangle _{n}=\frac{1}{t_{n}}\sum_{i=1}^{n}\mu_{i}\Delta t_{i},
\end{equation}
can be written as
\begin{equation}
R_{n}=e^{\left(\left\langle \mu\right\rangle _{n}-\left\langle \lambda\right\rangle _{n}\right)t_{n}}.
\end{equation}

Clearly, we have two different regimes as $n\rightarrow\infty$. If
$\left\langle \mu\right\rangle _{n}>\left\langle \lambda\right\rangle _{n}$,
then $R_{n}$ grows without limit and $\rho_{n}$ approaches zero,
indicating an inactive phase. On the other hand, if $\left\langle \mu\right\rangle _{n}<\left\langle \lambda\right\rangle _{n}$,
then $R_{n}$ approaches zero and $\rho_{n}$ remains finite, indicating
an active phase. Thus, the limiting case $\left\langle \mu\right\rangle _{n}=\left\langle \lambda\right\rangle _{n}$
signals the critical point. The behavior of the system exactly at
the critical point is governed by the fluctuations in the rates $\lambda_{n}$
and $\mu_{n}$, which depend on the precise way in which they are
chosen. 

For simplicity and without loss of generality, from now on we assume
a constant duration of each time interval, $\Delta t_{n}\equiv\Delta t$,
and set $\mu_{A}=\mu_{B}=\mu$. Thus, $\left\langle \mu\right\rangle _{n}=\mu$
and 
\begin{equation}
\left\langle \lambda\right\rangle _{N}=\frac{1}{N}\sum_{i=1}^{N}\lambda_{i}=\frac{N_{A}}{N}\lambda_{A}+\frac{N_{B}}{N}\lambda_{B},
\end{equation}
in which $N_{A}$ and $N_{B}$ are the numbers of letters $A$ and
$B$ in the generalized Fibonacci sequence of length $N$. The fraction
of letters in the infinite word (see Appendix~\ref{app:fractions})
are 
\[
x_{A}\equiv\lim_{N\rightarrow\infty}\frac{N_{A}}{N}=\zeta_{+}^{-1},
\]
 and $x_{B}\equiv1-x_{A}$, with 
\begin{equation}
\zeta_{\pm}=\frac{1\pm\sqrt{1+4k}}{2}.\label{eq:zetapm}
\end{equation}

The critical point $\lim_{N\rightarrow\infty}\left\langle \lambda\right\rangle _{N}=\lim_{N\rightarrow\infty}\left\langle \mu\right\rangle _{N}$
can, therefore, be recast as
\begin{equation}
x_{A}\lambda_{A}+x_{B}\lambda_{B}=\mu.\label{eq:mfcrit}
\end{equation}
 Assuming $\lambda_{A}<\lambda_{B}$, it is clear from Eq.~(\ref{eq:mfcrit})
that at the critical point we must have $\lambda_{A}<\mu<\lambda_{B}$.
Thus, sufficiently close to the critical point, $\rho\left(t\right)$
will decrease or increase during the $n$th time interval depending
on whether $\lambda_{n}=\lambda_{A}$ or $\lambda_{n}=\lambda_{B}$.
A plot of $\rho_{n}$ vs $n$ has the form illustrated in Fig.~\ref{fig:ilustrarenorma}
for $k=2$. The regions in which $\rho$ increases have a duration
equal to $k$ time intervals, while the regions in which $\rho$ decreases
last either $1$ or $k+1$ time intervals. 

\section{RG treatment\label{sec:RG}}

We are interested in describing the asymptotic behavior close to the
critical point. Since $\rho\left(t\right)\ll1$ in that case, then
we can disregard the term $s_{n}$ in Eq.~(\ref{eq:logrr}). This
is very helpful because only the knowledge of $\left\{ r_{n}\right\} $
determines completely the critical behavior of the system. In log-variables,
Eq.~(\ref{eq:logrr}) become an ``aperiodic'' walk (instead of a
random walk) where the steps of the walker is $\ln r_{n}$. Our task
now is to determine the properties of this walker. 

It is convenient to group consecutive intervals having the same parameters
$\left\{ \lambda_{n}\right\} =\lambda$ into a single interval with
that parameter $\lambda$ and larger duration as depicted in Fig.~\ref{fig:ilustrarenorm}.
Therefore, instead of considering $r_{n}$ equal to $r_{B}=e^{\left(\mu-\lambda_{B}\right)\Delta t}$
or $r_{A}=e^{\left(\mu-\lambda_{A}\right)\Delta t}$ in Eq.~(\ref{eq:logrr}),
we need to deal with $r_{n}$ being equal to $r_{0}^{+}=r_{B}^{k}$,
$r_{0}^{-}=r_{A}$, and $r_{0}^{--}=r_{A}^{k+1}$. This is because
intervals in which $\lambda_{n}=\lambda_{B}$ only appear in a sequence
of $k$ $B$-intervals in a row. On the other hand, the $A$-intervals
either appear as a single one, or in a sequence of $k+1$ intervals
in a row. In addition, we have to consider non-uniform time intervals
$\Delta t_{0}^{+}=k\Delta t$, $\Delta t_{0}^{-}=\Delta t$, and $\Delta t_{0}^{++}=\left(k+1\right)\Delta t$,
respectively. 

\begin{figure*}
\begin{centering}
\subfloat[\label{fig:schemea}]{\begin{centering}
\includegraphics[clip,width=0.9\textwidth]{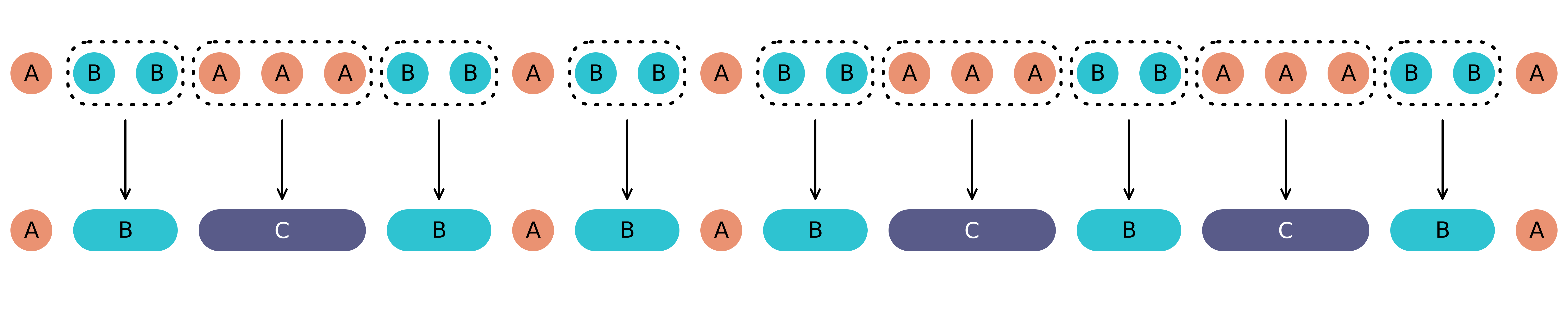}
\par\end{centering}

}
\par\end{centering}
\begin{centering}
\subfloat[\label{fig:schemeb}]{\centering{}\includegraphics[clip,width=0.9\textwidth]{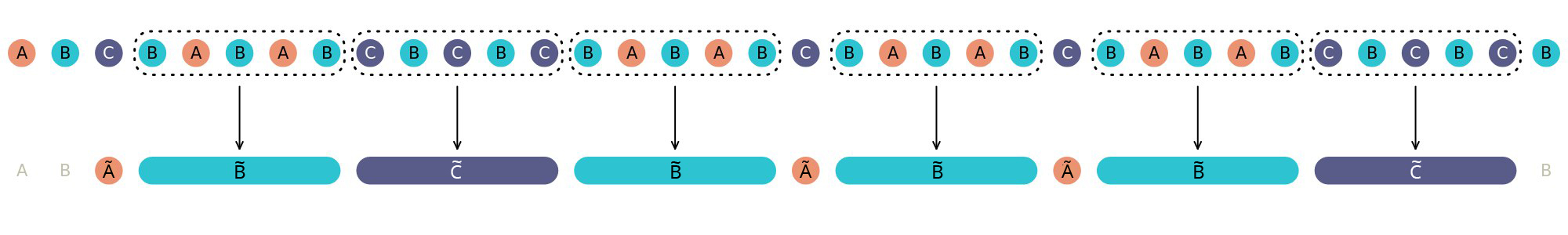}}\caption{\label{fig:esquema-renorm}Pictorial representation of the RG scheme
for $k=2$. (a) The initial grouping stage gives rise to three types
of effective time intervals, labeled $A$, $B$ and $C$ and characterized
by different effective parameters (see main text). The density of
active/infected sites increases in $B$-type intervals and decreases
in $A$- and $C$-type intervals. (b) Apart from boundary defects,
effective intervals generated in the next stage, labeled by $\tilde{A}$,
$\tilde{B}$ and $\tilde{C}$, follow the same sequence as the time
intervals $A$, $B$ and $C$.}
\par\end{centering}
\end{figure*}

In sum, each effective time interval in the regrouped system is characterized
by a pair of effective parameters $\left(\tilde{r},\tilde{\Delta t}\right)$
given by $\left(r_{0}^{++},\Delta t_{0}^{++}\right)$, $\left(r_{0}^{-},\Delta t_{0}^{-}\right)$,
or $\left(r_{0}^{--},\Delta t_{0}^{--}\right)$ as shown in Fig.~\ref{fig:ilustrac}.
As explicitly shown in Fig.~\ref{fig:schemea}, there are three types
($A$, $B$, and $C$) of intervals to consider.

The reason of the superscripts ``$\pm$'' is because we are assuming
that $\lambda_{A}<\mu<\lambda_{B}$, so that $+$ ($-$) means an
interval in which $\rho$ increases (decreases): $r_{0}^{++}<1<r_{0}^{-}<r_{0}^{--}.$
The reason for the subscript ``0'' is to call attention that these
are the bare values. As will become clear below, renormalized values
acquire a subscript $j$ denoting the number of times it was renormalized.

Following Ref.~\citealp{Vojta2015}, we now formulate an RG treatment
to iteratively determine the set of effective parameters $\tilde{r}$
and $\tilde{\Delta t}$ describing the long-time behavior of $\rho\left(t\right)\ll1$.
In the initial stage of the RG treatment, this set corresponds to
$\left\{ \left(r_{0}^{++},\Delta t_{0}^{++}\right),\left(r_{0}^{-},\Delta t_{0}^{-}\right),\left(r_{0}^{--},\Delta t_{0}^{--}\right)\right\} $.
At any given stage, we identify the effective parameter $\ln\tilde{r}$
closest to $0$, which characterizes those effective time intervals
during which the density varies the least, and use Eq.~(\ref{eq:logrr})
to eliminate all those time intervals, generating a new configuration
of effective time intervals and defining a new stage of the RG scheme,
see Fig.~\ref{fig:schemeb}. Importantly, as we checked numerically,
in each stage $j$ of this decimation procedure the temporal sequence
of effective time intervals is always the same (except for minor boundary
effects). Precisely, the effective parameters in the $j$th stage
are 
\begin{equation}
\left(\begin{array}{c}
\ln r_{j}^{++}\\
\ln r_{j}^{-}\\
\ln r_{j}^{--}
\end{array}\right)=\mathbf{M}\left(\begin{array}{c}
\ln r_{j-1}^{++}\\
\ln r_{j-1}^{-}\\
\ln r_{j-1}^{--}
\end{array}\right)=\mathbf{M}^{j}\left(\begin{array}{c}
\ln r_{0}^{++}\\
\ln r_{0}^{-}\\
\ln r_{0}^{--}
\end{array}\right),\label{eq:lnaj}
\end{equation}
\begin{equation}
\left(\begin{array}{c}
\Delta t_{j}^{++}\\
\Delta t_{j}^{-}\\
\Delta t_{j}^{--}
\end{array}\right)=\mathbf{M}\left(\begin{array}{c}
\Delta t_{j-1}^{++}\\
\Delta t_{j-1}^{-}\\
\Delta t_{j-1}^{--}
\end{array}\right)=\mathbf{M}^{j}\left(\begin{array}{c}
\ln r_{0}^{++}\\
\ln r_{0}^{-}\\
\ln r_{0}^{--}
\end{array}\right),\label{eq:Dtj}
\end{equation}
with
\begin{equation}
\mathbf{M}=\left(\begin{array}{ccc}
k+1 & k & 0\\
0 & 0 & 1\\
k & 0 & k+1
\end{array}\right),\label{eq:matrixM}
\end{equation}
as long as, in each stage, the constraint 
\begin{equation}
r_{j}^{++}<1<r_{j}^{-}<r_{j}^{--}\label{eq:ineqcond}
\end{equation}
is fulfilled. This always happens at the critical point, but slightly
off criticality it eventually fails as discussed later. 

As shown in Appendix~\ref{app:diag}, 
\begin{equation}
\left(\begin{array}{c}
\ln r_{j}^{++}\\
\ln r_{j}^{-}\\
\ln r_{j}^{--}
\end{array}\right)=\left(\begin{array}{c}
\eta_{0}+\eta_{0}^{-}\Xi_{-}^{j}+\eta_{0}^{+}\Xi_{+}^{j}\\
\eta_{1}+\eta_{1}^{-}\Xi_{-}^{j}+\eta_{1}^{+}\Xi_{+}^{j}\\
\eta_{2}+\eta_{2}^{-}\Xi_{-}^{j}+\eta_{2}^{+}\Xi_{+}^{j}
\end{array}\right)\label{eq:lnaj-1}
\end{equation}
and
\begin{equation}
\left(\begin{array}{c}
\Delta t_{j}^{++}\\
\Delta t_{j}^{-}\\
\Delta t_{j}^{--}
\end{array}\right)=\left(\begin{array}{c}
\tau_{0}+\tau_{0}^{-}\Xi_{-}^{j}+\tau_{0}^{+}\Xi_{+}^{j}\\
\tau_{1}+\tau_{1}^{-}\Xi_{-}^{j}+\tau_{1}^{+}\Xi_{+}^{j}\\
\tau_{2}+\tau_{2}^{-}\Xi_{-}^{j}+\tau_{2}^{+}\Xi_{+}^{j}
\end{array}\right),\label{eq:Dtj-1}
\end{equation}
 where 
\begin{equation}
\Xi_{\pm}=\zeta_{\pm}+k=\zeta_{\pm}^{2}\label{eq:Xi_pm}
\end{equation}
are two of the eigenvalues of $\mathbf{M}$ (the remaining one equals
to $1$), and $\zeta_{\pm}$ is given in Eq.~(\ref{eq:zetapm}).
Expressions for the coefficients $\eta_{i}^{x}$ and $\tau_{i}^{x}$
are also presented in Appendix~\ref{app:diag}. 

Since $\Xi_{+}>\Xi_{-}>0$, the long-time behavior is, thus, governed
by the coefficients of $\Xi_{+}^{j}$, namely $\eta_{i}^{+}$ and
$\tau_{i}^{+}$, $i\in\left\{ 0,1,2\right\} $. This is true only
away from criticality since $\eta_{i}^{+}=\gamma_{i}^{+}\left[\mu-\left(x_{A}\lambda_{A}+x_{B}\lambda_{B}\right)\right],$
with $\gamma_{i}^{+}>0$ a $k$-dependent constant (see Appendix~\ref{app:diag}),
i.e., $\eta_{i}^{+}$ is proportional to the distance to criticality
Eq.~(\ref{eq:mfcrit}). Therefore, in the active phase $\eta_{i}^{+}<0$,
$r_{j}^{++}$, $r_{j}^{-}$ and $r_{j}^{--}$ become smaller and smaller
as the RG scheme is iterated {[}see Eq.~(\ref{eq:lnaj-1}){]} and,
eventually, $r_{j}^{--}$ becomes smaller than $1$. At this stage,
labeled $j=j^{*}$, the constraint (\ref{eq:ineqcond}) is no fulfilled
and the RG must be interrupted. The value of $j^{*}$ can be estimated
by solving the equation 
\begin{equation}
\ln r_{j^{*}}^{--}=\eta_{2}+\eta_{2}^{-}\Xi_{-}^{j^{*}}+\eta_{2}^{+}\Xi_{+}^{j^{*}}=0.
\end{equation}

Sufficiently close to criticality, we expect $j^{*}\gg1$. For $k=1$,
$\Xi_{-}<1$ and, thus, 
\begin{equation}
\Xi_{+}^{j^{*}}=-\frac{\eta_{2}}{\eta_{2}^{+}}\sim\epsilon^{-1}\quad\left(k=1\right),\label{eq:sxk1}
\end{equation}
 with $\epsilon\equiv\mu-x_{A}\lambda_{A}-x_{B}\lambda_{B}$ being
the distance from criticality. For $k=2$, $\Xi_{-}=1$ and, thus,
\begin{equation}
\Xi_{+}^{j^{*}}=-\frac{\eta_{2}+\eta_{2}^{-}}{\eta_{2}^{+}}\sim\epsilon^{-1}\quad\left(k=2\right).\label{eq:sxk2}
\end{equation}
 Finally for $k>2$, $\Xi_{-}>1$ and, thus, 
\begin{equation}
j^{*}\approx\frac{\ln\left(-\eta_{2}^{-}/\eta_{2}^{+}\right)}{\ln\left(\Xi_{+}/\Xi_{-}\right)}\sim\frac{\ln\left(1/\epsilon\right)}{\ln\left(\Xi_{+}/\Xi_{-}\right)}.\label{eq:sxk3}
\end{equation}

The next step of our reasoning is to realize that the quantity $\Delta t_{j^{*}}^{--}$
plays the role of a characteristic time scale (when the critical RG
flow breaks down), i.e., the correlation time $\xi_{t}\sim\Delta t_{j^{*}}^{--}$.
From Eq.~(\ref{eq:Dtj-1}), we then conclude that 
\begin{equation}
\xi_{t}\sim\Delta t_{s_{\epsilon}}^{--}\sim\Xi_{+}^{j^{*}}.\label{eq:xit-sx}
\end{equation}
 It is now clear the usefulness of Eqs.~(\ref{eq:sxk1})--(\ref{eq:sxk3}).
From Eq.~(\ref{eq:def-nu}), we find that $\nu_{\parallel}=1$ for
$k\leq2$, and $\nu_{\parallel}=\frac{\ln\Xi_{+}}{\ln\frac{\Xi_{+}}{\Xi_{-}}}=\frac{\ln\left(\frac{1+\sqrt{1+4k}}{2}\right)}{\ln\left(\frac{\sqrt{1+4k}+1}{\sqrt{1+4k}-1}\right)}$
for $k>2$. Using the fact that $\Xi_{\pm}=\zeta_{\pm}^{2}$ and invoking
the definition of the wandering exponent $\omega=\ln\left|\zeta_{-}\right|/\ln\zeta_{+}$
{[}see Eq.~(\ref{eq:omega}){]}, we conclude that 
\begin{equation}
\nu_{\parallel}=\max\left\{ 1,\frac{1}{1-\omega}\right\} .\label{eq:nu}
\end{equation}
 For $k<2$ the correlation time exponent takes the same value $\nu_{\parallel}=1$
as in the uniform limit $\lambda_{A}=\lambda_{B}$, as expected from
the criterion (\ref{eq:relev}). For the marginal case $k=2$, $\nu_{\parallel}$
also follows the clean value. In this case, the criterion cannot say
if aperiodic temporal disorder is a relevant perturbation or not.
Finally, for $k>2$, the criterion (\ref{eq:relev}) ensures that
the clean theory is relevant ($\omega_{k>2}>\omega_{c}$) and, thus,
a new universality class must take place. This is indeed the case
as the correlation time exponent acquire different values from the
clean theory.

We evaluated the critical exponent $\beta$ governing the behavior
of the density near criticality by numerical means (see Sec.~\ref{sec:Numerical-results}),
as we could not find a way around the difficulties in analytically
estimating the asymptotic density. To high accuracy, we find that
$\beta=1$ for all $k$. From Eq.~(\ref{eq:screl}), then 
\begin{equation}
\delta=\frac{1}{\nu_{\parallel}}=\min\left\{ 1,1-\omega\right\} .\label{eq:delta}
\end{equation}

We now compare our findings for deterministic aperiodic temporal disorder
with those for random disorder. In Ref.~\citealp{Wada2018} it was
shown that 
\begin{equation}
\nu_{\parallel}=\frac{2}{\gamma},\quad\delta=\frac{\gamma}{2},\quad\beta=1,\label{eq:exponents}
\end{equation}
 where $\gamma$ is the exponent of the power-law correlation between
disorder variables. As previously mentioned, $\gamma=2-2H$ where
$H$ is the so-called Hurst exponent, which measures the long-term
memory of a time series. To our purposes, the identification between
$H$ and $\omega$ follows from the following. The wandering exponent
quantifies how the variance of a given letter in a word of size $N$
grows with $N$. Precisely, see Appendix~\ref{app:fractions}, the
variance $\sim N^{2\omega}$. If this word were the time series of
correlated random variables, the variance would grow $\sim N^{2H}$
as this is the definition of the Hurst exponent.

In sum, by identifying the wandering exponent $\omega$ to $H$ in
our results (\ref{eq:nu}) and (\ref{eq:delta}), we then recover
Eq.~(\ref{eq:exponents}). This fascinating result allows us to pinpoint
the precise fluctuation governing the relevance of the disorder on
this nonequilibrium phase transition, regardless whether disorder
is of random correlated character or aperiodically deterministic.

\section{Numerical results\label{sec:Numerical-results}}

We numerically iterated Eq.~(\ref{eq:logrr}) for $k=1$, $k=2$,
$k=3$ and various choices of the parameters $\mu$, $\lambda_{A}$
and $\lambda_{B}$, focusing on the neighborhood of the critical point.
In order to make it easier to identify the asymptotic behavior, we
also performed averages of the results over many aperiodic samples
with the same number of time intervals. These samples are defined
by randomly choosing the initial time interval among the positions
of a very large generalized Fibonacci sequence. With this choice,
all samples are representative of the infinite sequence and, if the
calculation is performed up to a sufficiently large time, no two samples
are likely to be equal. We initialize all samples with the same nonzero
value of the density of infected sites, which, as we checked, has
no effect on the average long-time behavior.

Besides looking at the time dependence of the average density over
all samples, $\left\langle \rho\left(t\right)\right\rangle $, we
also analyzed the dynamical evolution of the critical noise, quantified
both by $\sigma_{\rho}$$\left(t\right)$, the standard deviation
of $\rho\left(t\right)$ at time $t$ for all available samples, and
by $\sigma_{\ln\rho}\left(t\right)$, the corresponding quantity for
$\left\langle \ln\rho\left(t\right)\right\rangle $. As shown below,
the ratios $\sigma_{\rho}\left(t\right)/\left\langle \rho\left(t\right)\right\rangle $
and $\sigma_{\ln\rho\left(t\right)}/\left\langle \ln\rho\left(t\right)\right\rangle $
offer insight on the asymptotic behavior under temporal disorder inducing
fluctuations characterized by different wandering exponents $\omega$. 

\subsection{Case $k=1$}

\begin{figure}[t]
\begin{centering}
\subfloat[\label{fig:k=00003D1-a}]{\begin{centering}
\includegraphics[clip,width=0.9\columnwidth]{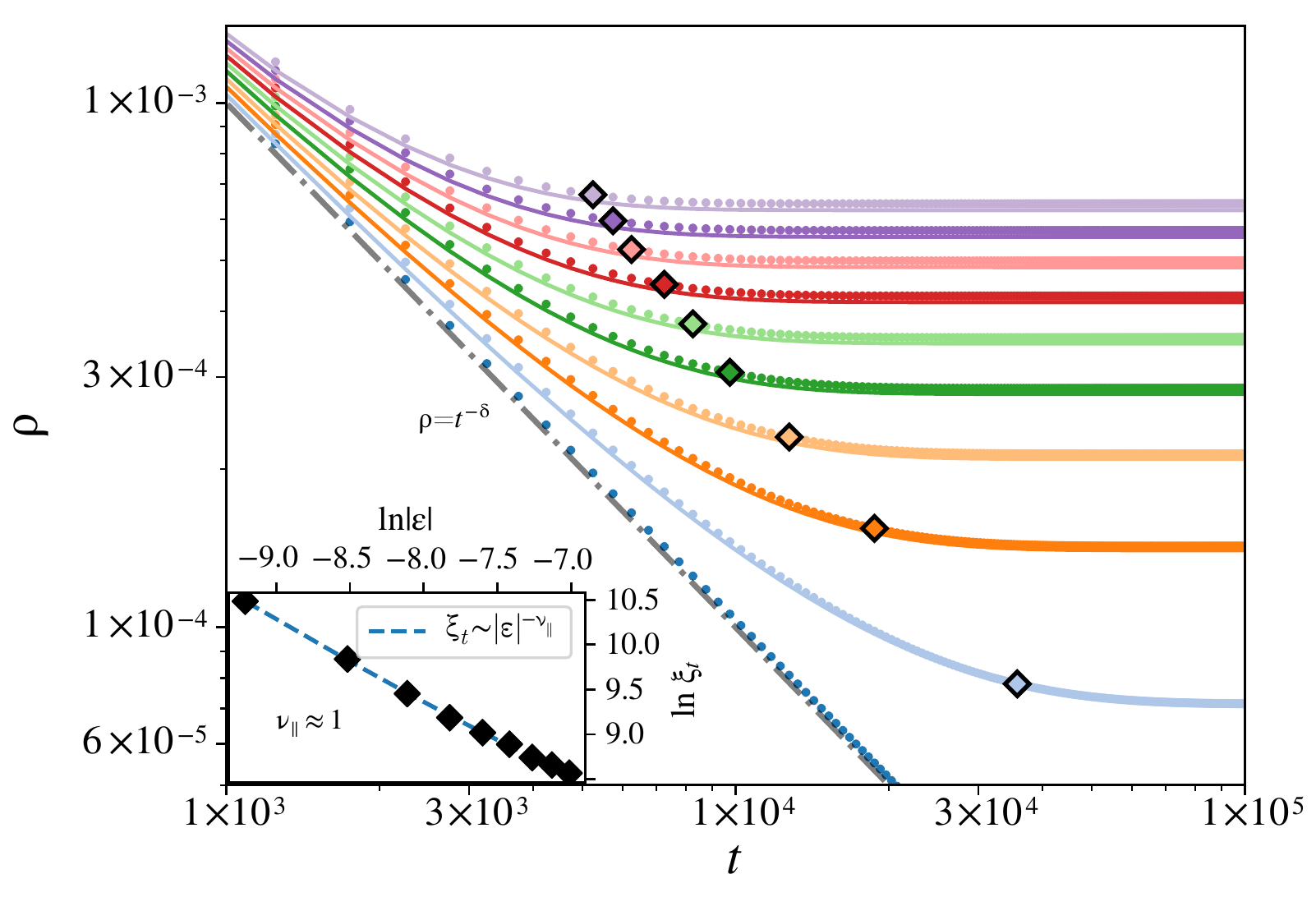}
\par\end{centering}
}
\par\end{centering}
\begin{centering}
\subfloat[\label{fig:k=00003D1-b}]{\centering{}\includegraphics[clip,width=0.9\columnwidth]{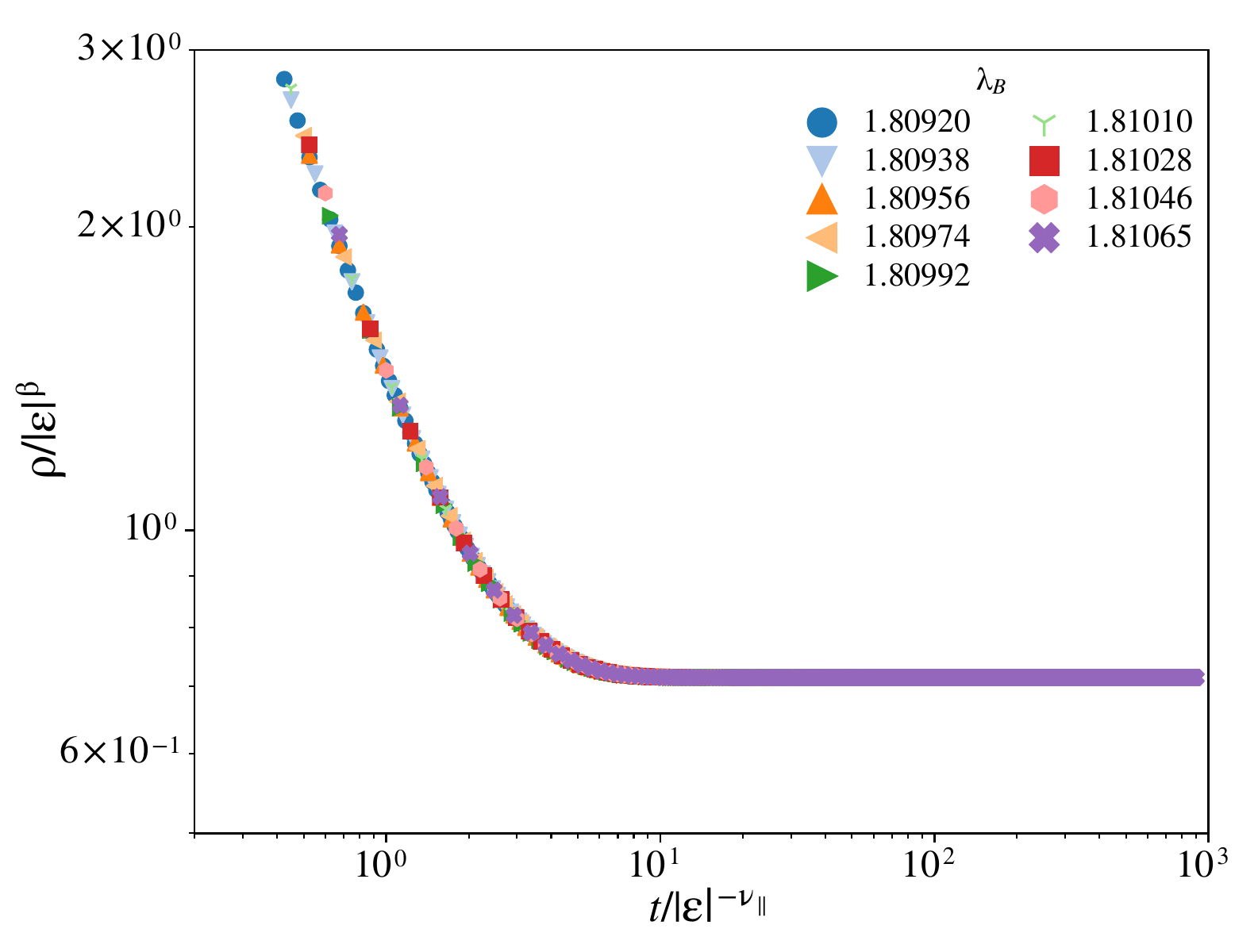}}
\par\end{centering}
\caption{\label{fig:dens_k=00003D1}(a) The main plot shows $\left\langle \rho\left(t\right)\right\rangle $
vs $t$ in the active phase, for $k=1$ with $\mu=1$, $\lambda_{A}=1/2$
and distances to criticality from $\epsilon=9\times10^{-4}$ to $\epsilon=0$,
top to bottom. Symbols indicate estimates for the characteristic time
$\xi_{t}$, whose log-log dependence on the distance to the critical
point is shown in the inset. The estimates come from determining,
for each curve, the time at which the average density reaches a value
10\% above its asymptotic value. (b) Rescaled plots of the average
density, showing data collapse following Eq. (\ref{eq:scform}). Here,
$\beta=\delta=\nu_{\parallel}=1$ are the values of the clean theory.}

\end{figure}
Plots of $\left\langle \rho\left(t\right)\right\rangle $ in the active
phase and at the critical point are shown in Fig.~\ref{fig:k=00003D1-a}.
It is clear that the behavior is quite similar to that of the uniform
limit, as illustrated by the fact that all curves closely follow those
obtained for a uniform system with the same average parameters as
the corresponding aperiodic system. At the critical point, the behavior
of $\left\langle \rho\left(t\right)\right\rangle $ is perfectly compatible
with a power law $t^{-\delta}$, with $\delta=\beta/\nu_{\parallel}=1$
as in the uniform model. In fact, as shown in Fig.~\ref{fig:k=00003D1-b},
all curves can be collapsed onto the same scaling form
\begin{equation}
\rho\left(t;\epsilon\right)=\left|\epsilon\right|^{\beta}f\left(t\left|\epsilon\right|^{\nu_{\parallel}}\right),\label{eq:scform}
\end{equation}
with $\beta=\nu_{\parallel}=1$, in which $f\left(t\left|\epsilon\right|^{\nu_{\parallel}}\right)$
is a scaling function taking a constant value if $t\gg\left|\epsilon\right|^{-\nu_{\parallel}}\sim\xi_{t}$.
For definiteness, we fixed both $\mu$ and $\lambda_{A}$ and use
$\lambda_{B}$ as a tuning parameter to cross the transition at the
critical value
\begin{equation}
\lambda_{B}^{*}=\frac{\mu\zeta_{+}-\lambda_{A}}{\zeta_{+}-1},\label{eq:lbcrit}
\end{equation}
 which is obtained from Eq.~(\ref{eq:mfcrit}). In that case, the
distance from criticality is defined as 
\begin{equation}
\epsilon\equiv1-\frac{\lambda_{B}}{\lambda_{B}^{*}}.
\end{equation}

\begin{figure}
\begin{centering}
\includegraphics[clip,width=1\columnwidth]{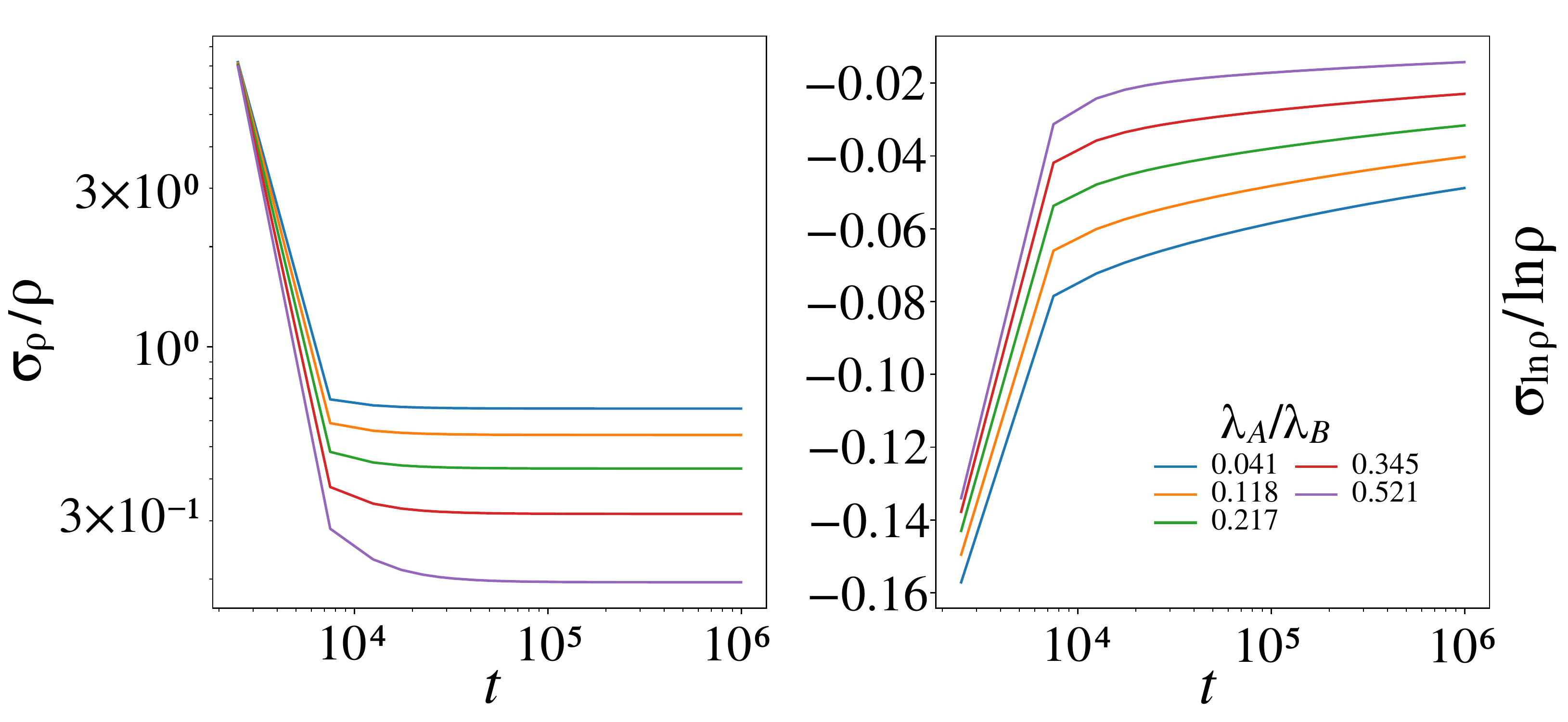}
\par\end{centering}
\caption{\label{fig:ratios_k=00003D1}Plots of linear (left) and logarithmic
(right) noise ratios at criticality for $k=1$ and different modulation
strengths $\lambda_{A}/\lambda_{B}$. For a given modulation strength,
the critical value of $\mu$ is determined by using Eq.~(\ref{eq:mfcrit}).}
\end{figure}
The behavior of the ratios $\sigma_{\rho}\left(t\right)/\left\langle \rho\left(t\right)\right\rangle $
and $\sigma_{\ln\rho\left(t\right)}/\left\langle \ln\rho\left(t\right)\right\rangle $
at criticality is shown in Fig.~\ref{fig:ratios_k=00003D1}. These
plots can be understood by noticing that for $k=1$ and at large times,
$\rho_{i}\left(t\right)\sim C_{i}t^{-1}$, in which $i$ labels a
given sample and all $C_{i}$ are approximately the same, given the
fact that fluctuations are small. Thus, denoting by $\sigma_{C}$
the standard deviation of the $C_{i}$, we have 
\begin{equation}
\left\langle \rho\left(t\right)\right\rangle \sim\frac{\left\langle C_{i}\right\rangle }{t},\quad\sigma_{\rho}\left(t\right)\sim\frac{\sigma_{C}}{t},
\end{equation}
so that
\begin{equation}
\frac{\sigma_{\rho}\left(t\right)}{\left\langle \rho\left(t\right)\right\rangle }\sim\frac{\left\langle C_{i}\right\rangle }{\sigma_{C}},
\end{equation}
and the ratio $\sigma_{\rho}\left(t\right)/\left\langle \rho\left(t\right)\right\rangle $
should approach a constant at large times. Likewise, denoting by $\sigma_{\ln C}$
the standard deviation of $\ln C_{i}$,
\[
\left\langle \ln\rho\left(t\right)\right\rangle =\left\langle \ln C_{i}\right\rangle -\ln t,\quad\sigma_{\ln\rho}=\sigma_{\ln C},
\]
so that
\[
\frac{\sigma_{\ln\rho}\left(t\right)}{\left\langle \ln\rho\left(t\right)\right\rangle }\sim\frac{\sigma_{\ln C}}{\left\langle \ln C_{i}\right\rangle -\ln t},
\]
and the ratio $\sigma_{\ln\rho\left(t\right)}/\left\langle \ln\rho\left(t\right)\right\rangle $
should exhibit a weak time dependence at large times. These expectations
are fully compatible with the numerical results shown in Fig.~\ref{fig:ratios_k=00003D1}.

\subsection{Case $k=2$}

\begin{figure}
\begin{centering}
\subfloat[\label{fig:2a}]{\centering{}\includegraphics[clip,width=0.9\columnwidth]{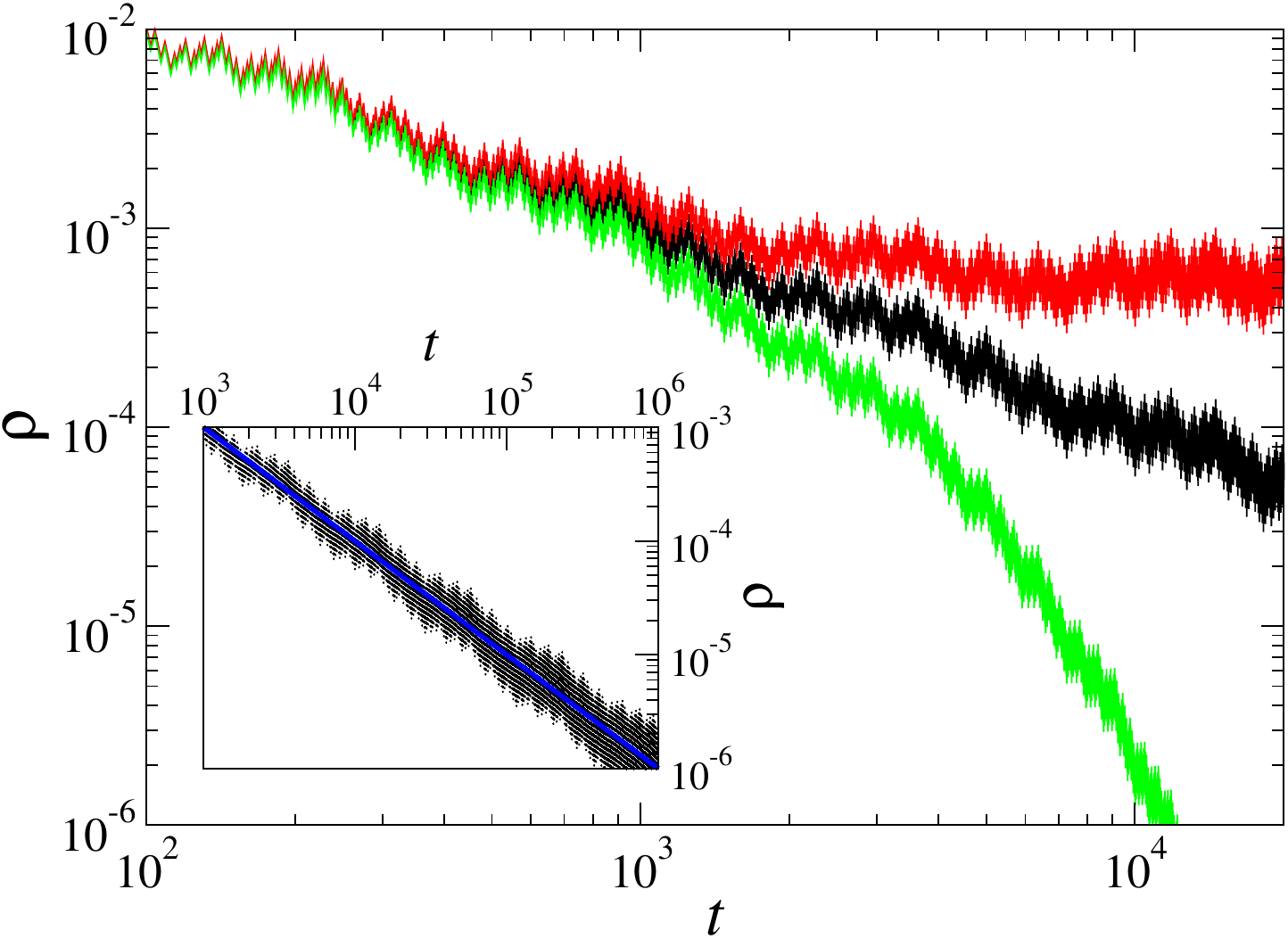}}
\par\end{centering}
\centering{}\subfloat[\label{fig:2b}]{\centering{}\includegraphics[clip,width=0.9\columnwidth]{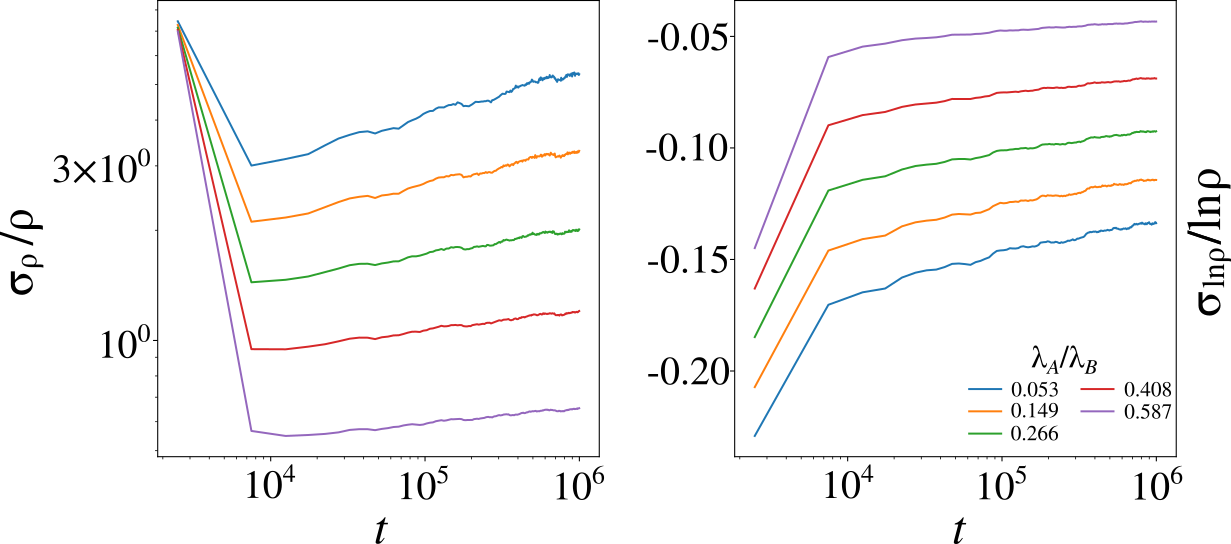}}\caption{\label{fig:dens_k=00003D2}(a) The main plot shows $\rho\left(t\right)$
vs $t$ for a single sample with $k=2$, $\mu=1$, $\lambda_{A}=9/10$
and distances to criticality given by $\epsilon=-10^{-3}$ (upper
red curve), $\epsilon=0$ (middle black curve), and $\epsilon=10^{-3}$
(lower green curve). The inset shows the long-time behavior of $\rho\left(t\right)$
at criticality, illustrating the slight increase in relative fluctuations
over time (for clarity, points are not connected by lines). The thick
blue curve is the function $\rho=1/t$. (b) Plots of linear (left)
and logarithmic (right) noise ratios at criticality for $k=2$ and
different modulation strengths $\lambda_{A}/\lambda_{B}$.}
\end{figure}
Now we analyze the marginal case $k=2$, which has a wandering exponent
$\omega=0$. Figure \ref{fig:2a} shows $\left\langle \rho\left(t\right)\right\rangle $
for different values of $\epsilon$. At the critical point the power-law
$\left\langle \rho\left(t\right)\right\rangle \sim t^{-\delta}$ is
still valid with $\delta=1$ as for $k=1$, but fluctuations are stronger.
This is also noticeable from the behavior of the ratio $\sigma_{\rho}\left(t\right)/\left\langle \rho\left(t\right)\right\rangle $
at criticality, shown in Fig.~\ref{fig:2b}. At long times, the ratio
no longer approaches a constant, but slightly increases as a power
law with an exponent that depends on the ratio $\lambda_{A}/\lambda_{B}$.
However, we cannot exclude the possibility of a logarithmic growth
with a ratio-dependent coefficient. This nonuniversality is characteristic
of marginal fluctuations. On the other hand, the ratio $\sigma_{\ln\rho\left(t\right)}/\left\langle \ln\rho\left(t\right)\right\rangle $
behaves similarly to the case $k=1$, approaching zero at long times.
This suggests that, as time increases, the relative width of the distribution
of $\rho_{i}\left(t\right)$ becomes larger, but that of $\ln\rho_{i}\left(t\right)$
becomes smaller.

\subsection{Case $k=3$}

\begin{figure}
\begin{centering}
\subfloat[\label{fig:3a}]{\begin{centering}
\includegraphics[clip,width=0.9\columnwidth]{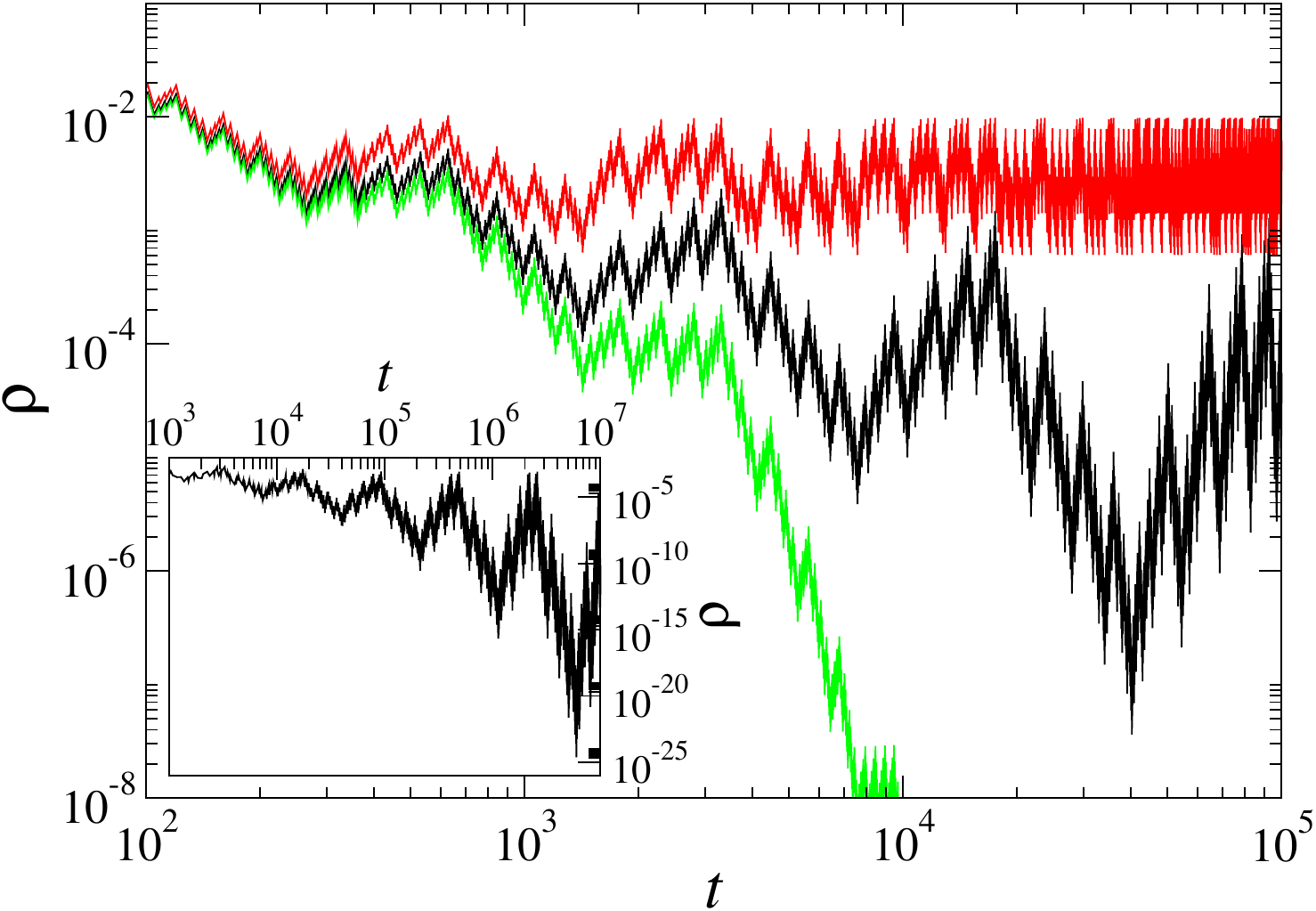}
\par\end{centering}
}
\par\end{centering}
\centering{}\subfloat[\label{fig:3b}]{\begin{centering}
\includegraphics[clip,width=0.9\columnwidth]{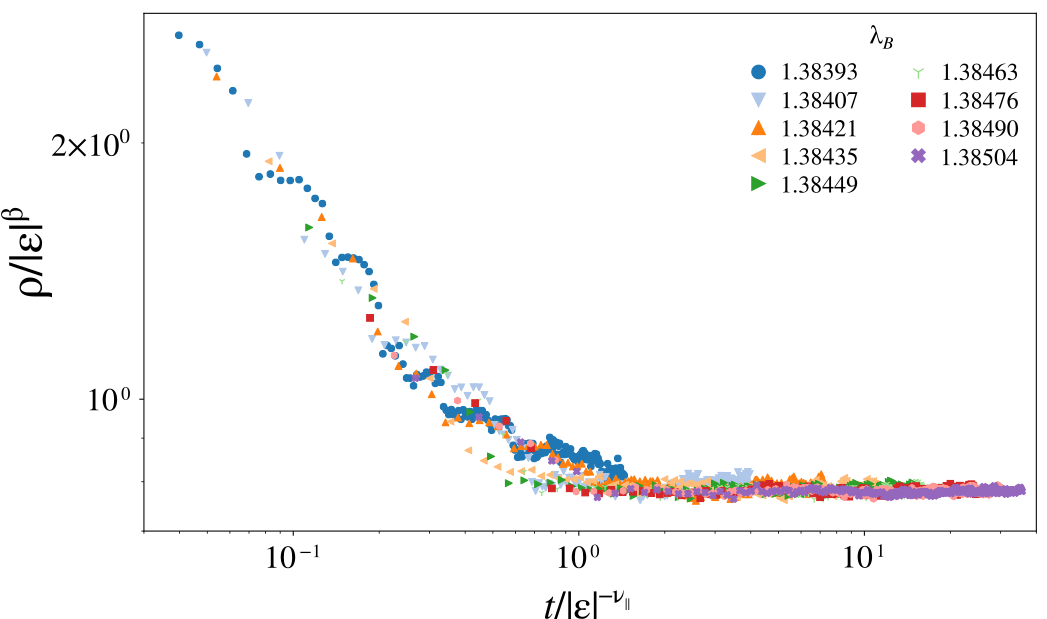}
\par\end{centering}
}\caption{\label{fig:dens_k=00003D3}(a) The main plot shows $\rho\left(t\right)$
vs $t$ for a single sample with $k=3$, $\mu=1$, $\lambda_{A}=9/10$
and distances to criticality given by $\epsilon=-5\times10^{-3}$
(upper red curve), $\epsilon=0$ (middle black curve), and $\epsilon=2\times10^{-3}$
(lower green curve). The inset shows the long-time behavior of $\rho\left(t\right)$
exactly at criticality, illustrating the strong increase in relative
fluctuations over time. (b) Rescaled plots of the average density
for $k=3$, with $\mu=1$ and $\lambda_{A}=1/2$, showing data collapse
following Eq.~(\ref{eq:scform}). Here, $\beta=1$ and $\nu_{\parallel}\approx1.46$
{[}see Eq.~(\ref{eq:nu}){]}.}
\end{figure}
Finally, we study the case $k=3$ in which aperiodic temporal disorder
is a relevant perturbation to the clean critical behavior. Here, the
wandering exponent is $\omega\approx0.317>\omega_{c}$ {[}see criterion
(\ref{eq:relev}){]}. For a single sample, density fluctuations increase
very strongly as a function of time at criticality, as shown in the
inset of Fig.~\ref{fig:3a}. When averaged over many samples, the
behavior is compatible with Eq.~(\ref{eq:scform}) with $\beta=1$
and $\nu_{\parallel}\approx1.46$, in agreement with the RG prediction
in Eq.~(\ref{eq:nu}), as shown in Fig.~\ref{fig:3b}.

\begin{figure}
\centering{}\includegraphics[clip,width=0.9\columnwidth]{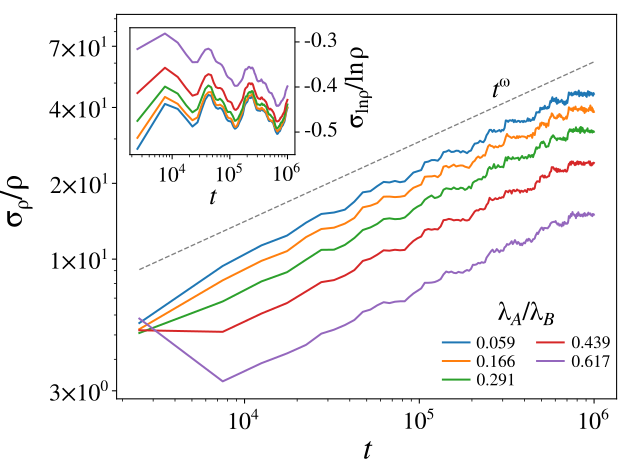}\caption{\label{fig:ratios_k=00003D3}Linear (main plot) and logarithmic (inset)
noise ratios at criticality for $k=3$ and different modulation strengths
$\lambda_{A}/\lambda_{B}$.}
\end{figure}
As for the ratios $\sigma_{\rho}\left(t\right)/\left\langle \rho\left(t\right)\right\rangle $
and $\sigma_{\ln\rho\left(t\right)}/\left\langle \ln\rho\left(t\right)\right\rangle $
at criticality, we can see from Fig.~\ref{fig:ratios_k=00003D3}
that $\sigma_{\rho}\left(t\right)/\left\langle \rho\left(t\right)\right\rangle $
follows a power-law with an exponent compatible with the wandering
exponent $\omega$ corresponding to $k=3$. (Although not shown, we
checked that the analogous behavior is also observed for other larger
values of $k$.) Furthermore, the ratio $\sigma_{\ln\rho\left(t\right)}/\left\langle \ln\rho\left(t\right)\right\rangle $
tends to oscillate around a constant at long times. This indicates
that, as time increases, the \emph{relative} width of the distribution
of $\rho_{i}\left(t\right)$ becomes larger, while that of $\ln\rho_{i}\left(t\right)$
remains constant. This should be compared with the behavior observed
under random temporal disorder~\citep{Vojta2015,Wada2018}, for which,
at criticality, $\sigma_{\ln\rho\left(t\right)}\sim\left|\ln\rho_{i}\right|\sim t^{(2-\gamma)/2}\sim t^{H}$,
also leading to a constant ratio $\sigma_{\ln\rho\left(t\right)}/\left\langle \ln\rho\left(t\right)\right\rangle $
at long times. Thus, our results indicate that, in the mean-field
limit, any wandering exponent $\omega>0$ leads to ``infinite-noise''
critical behavior at long times.

\section{Conclusions\label{sec:discussion}}

We have investigated the mean-field limit of the contact process in
the presence of deterministic aperiodic temporal disorder induced
by generalized Fibonacci sequences. These sequences have fluctuations
which grows with time as $\sim t^{\omega}$, with a wandering exponent
$\omega$ that depends of the parameter $k$ of the generalized sequences.
More importantly, the value of $\omega$ can be tuned such that aperiodic
temporal disorder can be a irrelevan, a marginal, or a relevant perturbation
to the clean critical behavior. For $\omega<\omega_{c}$ ($k<2$),
the long-time scaling behavior of the clean model remains unaltered,
with relative density fluctuations decreasing over time. For $\omega=\omega_{c}$
($k=2$), the aperiodic disorder induces density fluctuations which
grow slightly over time, but the critical exponents remain unaltered.
Finally, for $\omega>\omega_{c}$ ($k>2$), as in the case of random
temporal disorder, the long-time behavior is dominated by diverging
density fluctuations, and the critical behavior of the system is in
the so-called ``infinite-noise'' universality class. 

Nevertheless, contrary to the random case, aperiodic temporal disorder
does not give rise to active (temporal) Griffiths phases. In the random
case, these phases exist due to long (and rare) incursions of the
system in the inactive phase, even though the system is in the active
phase. These incursions allow the density to fall below any threshold
value associated with the inverse population size. Thus, the system
may reach the absorbing state even in the limit of an arbitrarily
large population. The underlying inflation symmetry of the generalized
Fibonacci sequences does not allow the formation of those long rare
regions. However, sufficiently close to criticality, finite regions
give rise to large fluctuations of the density at long times. This
is illustrated in the upper red curve of the main plot in Fig.~\ref{fig:3a}.

The long-time behavior described above is in full agreement with the
generalized criterion stated in Eq.~(\ref{eq:relev}). Such criterion
can be further tested for the contact process in finite dimensions,
as well as for other nonequilibrium models~\citep{Solano2016,Barghathi2017,Wada2021}.
It would also be interesting to investigate the effect of aperiodic
temporal disorder on systems exhibiting first-order nonequilibrium
phase transitions~\citep{Oliveira2016,Fiore2018,Encinas2021}.\bigskip{}

\begin{acknowledgments}
This work was supported by the Brazilian agencies CNPq and FAPESP.
A. P. V. acknowledges financial support from INCT/FCx. J.A.H. thanks
IIT Madras for a visiting position under the IoE program which facilitated
the completion of this research work.
\end{acknowledgments}

\appendix

\section{Properties of the generalized Fibonacci sequence\label{app:fractions}}

For the generalized Fibonacci sequence defined by the substitution
rule $A\rightarrow AB^{k}$ and $B\rightarrow A$, the numbers $N_{A}^{\left(j\right)}$
and $N_{B}^{\left(j\right)}$ of letters $A$ and $B$ in the finite
sequence obtained after $j$ iterations of the rule are given by the
matrix equation
\begin{equation}
\left(\begin{array}{c}
N_{A}^{\left(j\right)}\\
N_{B}^{\left(j\right)}
\end{array}\right)=\boldsymbol{\Omega}^{j}\left(\begin{array}{c}
1\\
0
\end{array}\right),
\end{equation}
in which we assume that the sequence is built starting from a single
letter $A$ and $\boldsymbol{\Omega}$ is the substitution matrix
\begin{equation}
\boldsymbol{\Omega}=\left(\begin{array}{cc}
1 & 1\\
k & 0
\end{array}\right).
\end{equation}
Diagonalizing $\boldsymbol{\Omega}$, we can write
\begin{equation}
\boldsymbol{\Omega}=\mathbf{U}\left(\begin{array}{cc}
\zeta_{+} & 0\\
0 & \zeta_{-}
\end{array}\right)\mathbf{U}^{-1},\qquad\mathbf{U}=\left(\begin{array}{cc}
\zeta_{-}/k & \zeta_{+}/k\\
1 & 1
\end{array}\right),
\end{equation}
with 
\begin{equation}
\zeta_{\pm}=\frac{1\pm\sqrt{1+4k}}{2},
\end{equation}
so that
\begin{equation}
\boldsymbol{\Omega}^{j}=\mathbf{U}\left(\begin{array}{cc}
\zeta_{+}^{j} & 0\\
0 & \zeta_{-}^{j}
\end{array}\right)\mathbf{U}^{-1},
\end{equation}
leading to
\begin{equation}
N_{A}^{\left(j\right)}=\frac{\zeta_{+}^{j+1}-\zeta_{-}^{j+1}}{\sqrt{4k+1}},\quad N_{B}^{\left(j\right)}=k\frac{\zeta_{+}^{j}-\zeta_{-}^{j}}{\sqrt{4k+1}}.
\end{equation}

Taking into account that $\zeta_{+}>\left|\zeta_{-}\right|$, the
asymptotic fractions of letters $A$ and $B$ are, respectively, 
\begin{equation}
x_{A}=\lim_{j\rightarrow\infty}\frac{N_{A}^{\left(j\right)}}{N_{j}}=\frac{1}{\zeta_{+}}\label{eq:xA}
\end{equation}
 and 
\begin{equation}
x_{B}=\lim_{j\rightarrow\infty}\frac{N_{B}^{\left(j\right)}}{N_{j}}=1-\frac{1}{\zeta_{+}},\label{eq:xB}
\end{equation}
 and thus,
\begin{equation}
N_{j}=N_{A}^{\left(j\right)}+N_{B}^{\left(j\right)}\sim\zeta_{+}^{j+2}.
\end{equation}

On the other hand, the fluctuations in the number of letters with
respect to the asymptotic expectation values, gauged by
\begin{equation}
G_{j}=\left|N_{A}^{\left(j\right)}-x_{A}N_{j}\right|,
\end{equation}
 are governed by
\begin{equation}
G_{j}\approx\frac{1}{\sqrt{4k+1}}\left|\zeta_{-}^{j}\left[\zeta_{-}-x_{A}\left(\zeta_{-}-k\right)\right]\right|\propto\left|\zeta_{-}^{j}\right|\propto N_{j}^{\omega},\label{eq:Gj}
\end{equation}
 which defines the wandering exponent
\begin{equation}
\omega=\frac{\ln\left|\zeta_{-}\right|}{\ln\zeta_{+}}.\label{eq:omega}
\end{equation}
 If $\omega<0$, the geometrical fluctuations get smaller as the sequence
gets larger, and at long times the behavior should recover that of
the uniform limit. On the other hand, if $\omega>0$, fluctuations
become larger and larger. The case $\omega=0$ is marginal and may
give rise to nonuniversal behavior. For the generalized Fibonacci
sequence, we have $\omega=-1<0$ for $k=1$, $\omega=0$ for $k=2$,
and $\omega\approx0.317>0$ for $k\geq3$. 

\section{Diagonalizing the matrix $\mathbf{M}$\label{app:diag}}

The matrix $\mathbf{M}$ in Eq.~(\ref{eq:matrixM}) can be written
as
\begin{equation}
\mathbf{M}=\mathbf{V}\left(\begin{array}{ccc}
1 & 0 & 0\\
0 & \Xi_{-} & 0\\
0 & 0 & \Xi_{+}
\end{array}\right)\mathbf{V}^{-1},
\end{equation}
with $\Xi_{\pm}$ given by Eq.~(\ref{eq:Xi_pm}) and
\begin{equation}
\mathbf{V}=\left(\begin{array}{ccc}
-1 & -\zeta_{+}/k & -\zeta_{-}/k\\
1 & \Xi_{+}/k^{2} & \Xi_{-}/k^{2}\\
1 & 1 & 1
\end{array}\right).
\end{equation}
Therefore, 
\begin{equation}
\mathbf{M}^{j}=\mathbf{V}\left(\begin{array}{ccc}
1 & 0 & 0\\
0 & \Xi_{-}^{j} & 0\\
0 & 0 & \Xi_{+}^{j}
\end{array}\right)\mathbf{V}^{-1}.
\end{equation}

Using 
\begin{equation}
\left(\begin{array}{c}
\ln r_{0}^{++}\\
\ln r_{0}^{-}\\
\ln r_{0}^{--}
\end{array}\right)=\left(\begin{array}{c}
k\left(\mu-\lambda_{B}\right)\\
\mu-\lambda_{A}\\
\left(k+1\right)\left(\mu-\lambda_{A}\right)
\end{array}\right)
\end{equation}
and 
\begin{equation}
\left(\begin{array}{c}
\Delta t_{0}^{++}\\
\Delta t_{0}^{-}\\
\Delta t_{0}^{--}
\end{array}\right)=\left(\begin{array}{c}
k\\
1\\
k+1
\end{array}\right)\Delta t
\end{equation}
in Eqs.~(\ref{eq:lnaj}) and (\ref{eq:Dtj}), we obtain Eqs.~(\ref{eq:lnaj-1})
and (\ref{eq:Dtj-1}) with
\begin{equation}
\eta_{0}=\frac{\left(\mu-\lambda_{A}\right)-k\left(\mu-\lambda_{B}\right)}{k-2}=-\eta_{1}=-\eta_{2},
\end{equation}
\begin{equation}
\eta_{0}^{\pm}=\mp\Delta\left[\left(\Xi_{\pm}-k^{2}\right)\left(\mu-\lambda_{A}\right)+k\left(\zeta_{\pm}+k\zeta_{\mp}\right)\left(\mu-\lambda_{B}\right)\right],
\end{equation}
\begin{equation}
\eta_{1}^{\pm}=\mp\Delta\left[\left(\zeta_{\pm}+k\zeta_{\mp}\right)\left(\mu-\lambda_{A}\right)+k\left(\zeta_{\pm}-k\right)\left(\mu-\lambda_{B}\right)\right],
\end{equation}
\begin{eqnarray}
\eta_{2}^{\pm} & = & \mp\Delta\left\{ \left[\Xi_{\pm}-k\left(k-1\right)\zeta_{\pm}\right]\left(\mu-\lambda_{A}\right)\right.\nonumber \\
 &  & +\left.k\left(\Xi_{\pm}-k^{2}\right)\left(\mu-\lambda_{B}\right)\right\} ,
\end{eqnarray}
\begin{equation}
\tau_{0}=-\frac{k-1}{k-2}=-\tau_{1}=-\tau_{2},
\end{equation}
\begin{equation}
\tau_{0}^{\pm}=\mp\Delta\left[\left(\Xi_{\pm}-k^{2}\right)+k\left(\zeta_{\pm}+k\zeta_{\mp}\right)\right],
\end{equation}
\begin{equation}
\tau_{1}^{\pm}=\mp\Delta\left(\zeta_{\pm}-k\left(k-1\right)\right),
\end{equation}
\begin{equation}
\tau_{2}^{\pm}=\mp\Delta\left[\left(k+1\right)\Xi_{\pm}-k\left(k-1\right)\zeta_{\pm}-k^{3}\right],
\end{equation}
in which 
\begin{equation}
\Delta^{-1}=\left(k-2\right)\sqrt{1+4k}.
\end{equation}

It is interesting to notice that 
\begin{equation}
\eta_{i}^{\pm}=\gamma_{i}^{\pm}\left(\mu-\frac{1}{\zeta_{\pm}}\lambda_{A}-\left(1-\frac{1}{\zeta_{\pm}}\right)\lambda_{B}\right),
\end{equation}
 where $\gamma_{0}^{\pm}=\pm\Delta\left(\zeta_{\pm}\left(k^{2}-k-1\right)-k\right)$,
$\gamma_{1}^{\pm}=\pm\Delta\left(k\left(k-1\right)-\zeta_{\pm}\right)$,
and $\gamma_{2}^{\pm}=\pm\Delta\left(k\left(k^{2}-1\right)-\left(2k+1\right)\zeta_{\pm}-k^{2}\zeta_{\mp}\right)$.
It is easy to show that $\eta_{i}^{+}>0$ for $k\ge0$.

For $k=2$, $\eta_{i}$, $\eta_{i}^{-}$, $\tau_{i}$ and $\tau_{i}^{-}$
are divergent. However, the following useful quantities remain finite:
\begin{equation}
\lim_{k\rightarrow2}\left(\eta_{0}+\eta_{0}^{-}\right)=\frac{2}{9}\left(\mu+4\lambda_{A}-5\lambda_{B}\right),
\end{equation}

\begin{equation}
\lim_{k\rightarrow2}\left(\eta_{1}+\eta_{1}^{-}\right)=\frac{1}{9}\left(\mu-5\lambda_{A}-4\lambda_{B}\right),
\end{equation}
\begin{equation}
\lim_{k\rightarrow2}\left(\eta_{2}+\eta_{2}^{-}\right)=\frac{1}{9}\left(-5\mu-11\lambda_{A}+16\lambda_{B}\right),
\end{equation}
\begin{equation}
\lim_{k\rightarrow2}\eta_{0}^{+}=\frac{16}{9}\left[\mu-\frac{1}{2}\left(\lambda_{A}+\lambda_{B}\right)\right],
\end{equation}
\begin{equation}
\lim_{k\rightarrow2}\eta_{1}^{+}=\frac{8}{9}\left[\mu-\frac{1}{2}\left(\lambda_{A}+\lambda_{B}\right)\right],
\end{equation}
\begin{equation}
\lim_{k\rightarrow2}\eta_{2}^{+}=\frac{32}{9}\left[\mu-\frac{1}{2}\left(\lambda_{A}+\lambda_{B}\right)\right],
\end{equation}
 $\lim_{k\rightarrow2}\left(\tau_{0}+\tau_{0}^{-}\right)=\frac{2}{9},$
$\lim_{k\rightarrow2}\left(\tau_{1}+\tau_{1}^{-}\right)=\frac{1}{9},$
$\lim_{k\rightarrow2}\left(\tau_{2}+\tau_{2}^{-}\right)=-\frac{5}{9},$$\lim_{k\rightarrow2}\tau_{0}^{+}=\frac{16}{9},$
$\lim_{k\rightarrow2}\tau_{1}^{+}=\frac{8}{9},$ and $\lim_{k\rightarrow2}\tau_{2}^{+}=\frac{32}{9}$. 

\bibliographystyle{apsrev4-1}
\bibliography{ref}

\end{document}